\documentclass[amssymb,preprint,12pt,aps]{revtex4}
\usepackage{graphicx}
\begin{document}
\title{Quantum chaos in elementary quantum mechanics}
\author{Yu. Dabaghian}

\affiliation{
Keck Center for Integrative Neuroscience, Department of Physiology,
University of California San Francisco, 513 Parnassus Avenue, SHE-806
San Francisco, CA 94143-0444, e-mail: yura@phy.ucsf.edu}

\date{\today}
\begin{abstract}
We introduce an analytical solution to the one of the most familiar
problems from the elementary quantum mechanics textbooks. The following 
discussion provides simple illustrations to a number of general concepts 
of quantum chaology, along with some recent developments in the field and 
a historical perspective on the subject. 
%It is a short story of Chaos, de Broglie waves and quantum numbers.
\end{abstract} 
\maketitle

\section{Introduction}

Many aspects of the behavior of quantum systems can be understood and 
interpreted in terms of the dynamical characteristics of their classical 
counterparts. It is often possible to obtain {\em quantitatively} such 
important attributes of a quantum system as its spectra and even its wave 
function in terms of the appropriate classical quantities \cite{Gutzw,Cvit,Kay}.
Although the main objects of the classical dynamics, the dynamical trajectories, 
are not something that can be rigorously used in the context of quantum mechanics, 
they can often facilitate our understanding of quantum realm, by providing means 
for semiclassical interpretation.

As an illustration of this point, let us first look at the infinite square 
well problem. Let us consider a particle confined in the infinite square well
potential,
\begin{equation}
V(x)=\cases{ 0, &if $0< x< b$,
\cr \infty &if $x\leq 0$ or $x\geq L$.\cr}
\label{well}
\end{equation}
Classically, the dynamics of such particle is as simple as it can possibly be - 
the particle simply bounces periodically between the two walls. Geometrically its 
trajectory is a closed loop, which the particle traverses over and over again. The 
total length of the loop is twice the width of the well, ${\cal L}=2L$, and the 
period of motion is $T=2L/v$, where $v$ is the speed of the particle. As a result of 
the wall reflections, the momentum of the particle changes its sign, but its magnitude 
never changes.

Quantum mechanically, this system is nearly as simple to describe. The wave function 
$\psi(x)$ in this case is a combination of two plane waves,
\begin{equation}
\psi(x)=Ae^{ikx}+Be^{-ikx},
\label{wave}
\end{equation}
where $k=\sqrt{E}/\hbar$ is the particle's momentum, which must satisfy the boundary 
conditions $\psi(0)=\psi(L)=0$. This leads to the quantization condition
\begin{equation}
\sin(kL)=0,
\label{wellspec}
\end{equation}
and hence the quantum spectrum $k_{n}$ of this problem is given by
\begin{equation}
L\,k_{n}=\pi n.
\label{deb}
\end{equation}
Note, that the left hand side of this equation coincides with the classical
action integral,
\begin{equation}
S=\int_{\gamma}kdx=L\,k,
\label{act}
\end{equation}
taken along the trajectory $\gamma$ that connects the two turning points, and 
hence the condition (\ref{deb}) implies that the action takes only discrete 
values, $S_{n}=\pi n$.
Expressing the de Broglie wavelength $\lambda$ in terms of the momentum, 
$\lambda=2\pi/k$, the relationship (\ref{deb}) can be cast into the form
${\cal L}/\lambda_{n}=n$, which has a simple geometrical meaning. Apparently
the quantization condition (\ref{deb}) implies that the wave (\ref{wave}) 
must fit {\em geometrically} $n$ times onto the (only) classical periodic 
orbit that has the length ${\cal L}$. 
As the energy of the particle increases, its wavelength will become smaller 
and smaller, however the geometrical condition ${\cal{L}}/ \lambda_{n}=n$ always 
holds.

Such simple combination of physical and geometrical ideas were used by Bohr
and his school in around 1914 to provide the first explanations to one of the
most striking features of the quantum mechanical systems - the discrete nature 
of their energy spectra.
According to their views, the discreteness of the quantum spectra turns out 
to be essentially a consequence of the geometrical consistency of the wave 
mechanics.
Specifically, such ``wave-geometrical'' approach proved very successful
in early attempts to explain the experimentally observed emission-absorption 
spectrum of the Hydrogen atom. In fact, the results obtained in this way were 
exact, and that gave serious reasons to believe that these ideas were adequate 
to describe the quantum physics of the subatomic world.

However, the later attempts of Bohr, Sommerfeld, van Vleck, Born and others 
\cite {Bohr,VanVleck,Born} to continue the success of these ideas on more 
complicated atoms, failed completely. 
It was apparently impossible to explain even approximately the experimentally
observed spectrum of the Helium atom - the next simplest atom after Hydrogen 
in the periodic table of elements and hence the next best candidate for a 
successful treatment by means of such wave-geometrical quantization. Van Vleck 
wrote in 1922 \cite{VanVleck}:

\bigskip
{\em ``The conventional quantum theory of atomic structure does 
not appear able to account for the properties of even such a simple element 
as helium, and to escape from this dilemma some radical modification in the
ordinary conceptions of quantum theory or of the electron may be necessary''}.
\bigskip

What was be the nature of the difficulties that required such ``radical'' and
``conceptual''  modifications? In 1925 Max Born wrote \cite{Born}: 

\bigskip
{\em ``...the systematic application of the principles of the quantum theory... 
gives results in agreement with experiment only in those cases where the motion 
of a single electron is considered; it fails even in the treatment of the motion 
of the two electrons in the helium atom.

This is not surprising, for the principles used are not really consistent... A 
complete systematic transformation of the classical mechanics into a discontinuous 
mechanics is the goal towards which the quantum theory strives.''}
\bigskip

Seemingly one makes a very simple and natural move by trying to go from 
the {\em exactly solved} Hydrogen atom to the Helium, by adding just one 
more particle to the two body nucleus-electron system of the Hydrogen.
However, from the point of view of the contemporary classical mechanics,
this modest generalization turns an {\em integrable} two-body system
into a {\em non-integrable} three body system. 
Hence, in order to impose the ``geometrical consistency'' on the quantum 
waves that could describe the Helium atom in Bohr's approach, one would have 
to deal with the overwhelming geometrical complexity of its classical phase 
space. It means that all the familiar ``niceties'' of the generic chaotic 
systems, such as the exponential proliferation of the periodic orbits 
associated with extreme complexity of their shapes, would have to be taken
into account - something that could hardly had been done at Bohr's time.
Indeed, at the beginning of the last century chaos theory was just being 
developed in the works of Poincare, Lyapunov, Hadamard, Birkhoff and a few 
others \cite{Lyapunov,Poincare,Hadamard,Birkhoff}. 
Although the importance of the classical dynamical behavior for successful
quantization was realized by certain researchers \cite{Einstein},
it took about 60 years before the first semiclassical quantization procedure
for nonintegrable systems was outlined. As for the Helium atom, a semiclassical 
quantization scheme for it was proposed in 1992 \cite{Wintgen} using some recent 
developments of the quantum chaos theory \cite{Cvit}.

\section{Back to 1D potential wells}

Surprisingly, the difficulties of the semiclassical quantization of the
classically nonintegrable Helium atom system can be illustrated by means 
of elementary quantum mechanics. Below we shall consider a simple modification
of a completely transparent system (\ref{well}) that leads to a transition 
from classical integrability to a non-integrability and clarifies the 
corresponding outburst of complexity on quantum level.

Let us add a small step at the bottom of the potential (\ref{well}) and consider 
a point particle moving in the potential
\begin{equation}
V(x)=\cases{ 0, &if $0\leq x\leq b$, 
     \cr V, &if $b\leq x\leq L$,
     \cr \infty &if $x\leq 0$ or $x\geq L$,\cr}  
\label{pot}
\end{equation}
shown in Figure~1. 
%%%%%%%%%%%%%%%%%%%%%%%%%%%%%%%%%%%%%%%%%%%%%%%%%%%%%%%%%%%%%%%%%%%%%%
\begin{figure}
\begin{center}
\includegraphics{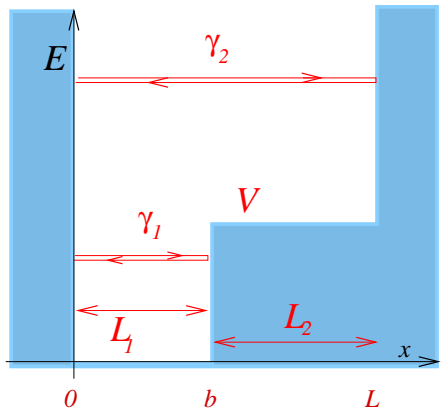}
\caption{\label{Fig.1} A simple modification of the infinite square well 
model - a square well with a potential step, along with the two apparent
classical trajectories $\gamma_{1}$ and $\gamma_{1}$.}
\end{center}
\end{figure}
%%%%%%%%%%%%%%%%%%%%%%%%%%%%%%%%%%%%%%%%%%%%%%%%%%%%%%%%%%%%%%%%%%%%%%
Seemingly the classical picture doesn't change much. It appears that if 
the energy of the particle is higher than $V$, the particle will 
oscillate between the points $x=0$ and $x=L$ just as before, and if its 
energy is below the potential step hight, it will oscillate between $x=0$ 
and $x=b$.
However, the reality turns out to be much more complicated than that.

Let us first give the quantum mechanical description of the problem. 
Suppose that the energy $E$ of the particle is above the potential height 
$V$. The momentum of the particle in the region $0\leq x\leq b$ is $k=\sqrt{E}$, 
and in the region $b\leq x\leq 1$ it is $\kappa=\sqrt{E-V}$, and its $\psi$ 
function consists of two parts,
\begin{equation}
\psi(x)=\cases{\sin(kx), &if $0\leq x\leq b$, 
\cr A\sin\left[\kappa\left(L-x\right)\right], &if $b\leq x\leq L$,\cr}  
\label{psi}
\end{equation}
which match continuously at $x=b$. 
%%%%%%%%%%%%%%%%%%%%%%%%%%%%%%%%%%%%%%%%%%%%%%%%%%%%%%%%%%%%%%%%%%%%%%
\begin{figure}
\begin{center}
\includegraphics{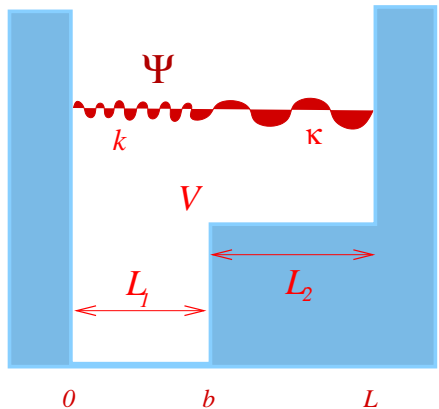} 
\caption{\label{Fig.2} A quantum particle in a square well with a step.}
\end{center}
\end{figure}
%%%%%%%%%%%%%%%%%%%%%%%%%%%%%%%%%%%%%%%%%%%%%%%%%%%%%%%%%%%%%%%%%%%%%%
From this continuity requirement one can determine the spectral equation for 
the spectrum of the problem,
\begin{equation}
\sin(L_{1}k+L_{2}\kappa)=r\sin\left[\left(L_{1}k-L_{2}\kappa\right)\right], \ 
\ 
\ E>V,
\label{specabove}
\end{equation}
where $L_{1}=b$ and $L_{2}=L-b$ are correspondingly the lengths of the left and 
the right sides of the well (\ref{pot}) and $r$ is the reflection coefficient
\begin{equation}
r=\frac{k-\kappa}{k+\kappa}.
\label{reflection}
\end{equation}
The $E<V$ case can be treated similarly and basically amounts to substituting
$\kappa\rightarrow i\varkappa$, $\varkappa=\sqrt{V-E}$, in the expressions 
(\ref{specabove}) and (\ref{reflection}). After extracting the complex phase, 
the spectral equation becomes
\begin{equation}
\sin(L_{1}k)=\frac{(-1)^{n+1} k\left(1-e^{-2L_{2}\varkappa}\right)}
{\sqrt{V\left(1+e^{-4L_{2}\varkappa}\right)+2(\varkappa^{2}-k^{2})e^{-2L_{2}
\varkappa}}}, \ \ \ E<V,
\label{specbelow}
\end{equation}
where $n=1$, $2$, ..., is the root index.

There is an important analogy between (\ref{specabove}), (\ref{specbelow}) and
(\ref{wellspec}) - the arguments of the sine functions in the left hand sides 
of these equations are the action lengths $S(E)$ of the classically available 
regions in the well (\ref{pot}) for $E>V$ and $E<V$ correspondingly,
\begin{equation}
S(k)=\int_{\gamma}kdx=
\cases{kL_{1}+\kappa L_{2}, &if $E>V$, 
\cr 
kL_{1} &if $E<V$,\cr}  
\label{potaction}
\end{equation}
which is a continuous and monotonically increasing function of the energy. 

On the other hand, it is interesting that unlike (\ref{wellspec}), the equations 
(\ref{specabove}) for $E>V$ and (\ref{specbelow}) for $E<V$ are {\em transcendental 
equations} - that is substantially more difficult to solve. In fact, in absence of 
any analytical ways of solving the equation (\ref{specabove}) and (\ref{specbelow}) 
explicitly, all the standard textbooks, e.g. \cite{Griffith,Flugge,Messiah,LL}, use 
graphical or numerical methods to approximate its roots, $k_{n}$.
%%%%%%%%%%%%%%%%%%%%%%%%%%%%%%%%%%%%%%%%%%%%%%%%%%%%%%%%%%%%%%%%%%%%%%
\begin{figure}
\begin{center}
\includegraphics{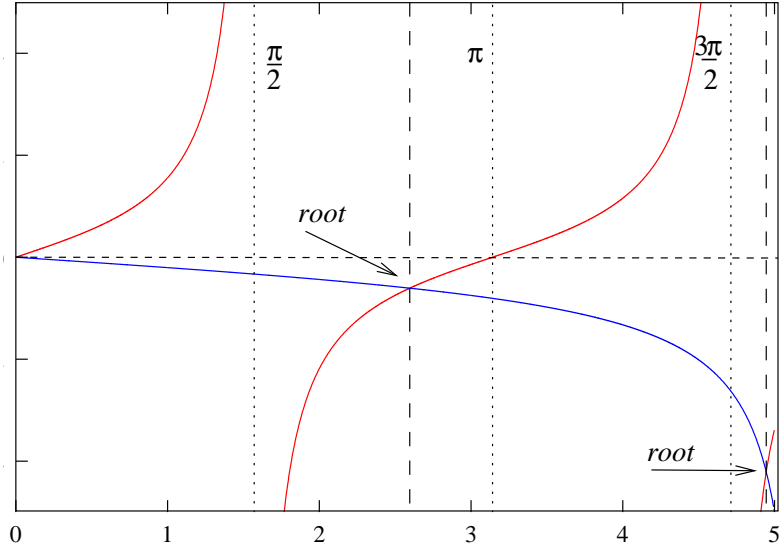}
\caption{\label{Fig.3} Typical appearance of the graphical methods currently used by 
the Quantum Mechanics textbooks. The roots are found as the intersection points of the 
graphs representing the right and the left hand sides of a suitable spectral equation, 
marked by the vertical dashed lines. For details see \cite{Griffith,Flugge,Messiah,LL}.}
\end{center}
\end{figure}
%%%%%%%%%%%%%%%%%%%%%%%%%%%%%%%%%%%%%%%%%%%%%%%%%%%%%%%%%%%%%%%%%%%%%%
Notably, there are also no simple ``semiclassical'' interpretations similar to 
(\ref{deb}) for the results of such approximations, which, after all, does not appear 
important for the purpose of obtaining the numerical values for the roots of the spectral 
equations (\ref{specabove}) and (\ref{specbelow}).
 
This is very similar to what happened to the Helium atom quantization problem - the 
``old quantum theory'' treatment became unnecessary once the formalism of the ``new'' 
quantum mechanics of Schr\"odinger and Heisenberg was established and could be used. 
The problem of obtaining the energy spectrum of the Helium (or in principle any other 
atom) was reduced to a {\em technical} problem of diagonalizing the Hamiltonian matrix 
$H_{nm}$ by means of all sorts of numerical techniques and approximations. 

Even without making any historical references, one can notice the obvious 
contrast in the level of complexity between the spectral equations (\ref{specabove}),
(\ref{specbelow}) and (\ref{wellspec}). What is the physical reason for it? 
Below we shall argue that the complexity of the spectral equations (\ref{specabove}), 
(\ref{specbelow}) has a deep physical meaning and can be understood from analyzing the 
classical motion of a particle in the potential well (\ref{pot}).

\section{The classical limit}

To start the discussion of the classical dynamics of a point particle in 
the potential (\ref{pot}), let us notice that the reflection coefficient 
(\ref{reflection}) in (\ref{specabove}) ($E>V$) does not depend on the 
Plank's constant $\hbar$,
\begin{equation}
r=\frac{1-\sqrt{1-V/E}}{1+\sqrt{1-V/E}}.
\label{one}
\end{equation}
Due to this circumstance this quantity (although obtained by purely quantum
mechanical means \cite{LL}) is in fact quite classical, because one does not 
need to refer to any quantum mechanical concepts to evaluate $r$.

This curious fact can be easily interpreted and understood from 
general principles \cite{RS1,RS2}.
Just as any other wave, the probability amplitude (\ref{wave}) gets 
reflected or diffracted when it encounters inhomogeneities on its way.
It is important that the scale of these inhomogeneities should not be 
too big compared to the wavelength of the wave, otherwise the wave 
will ``adjust'' to the smooth changes of the properties of the media.
It is fair to say that in order to induce reflections, the obstacle 
should appear somewhat ``abruptly'' in front of the wave, at a scale
$d$ smaller than the characteristic wavelength, $d\ll\lambda$.
In our case, since the potential step (\ref{pot}) is defined to be 
absolutely sharp, changing discontinuously at $x=b$ from $0$ to $V$, 
the quantum-mechanical wave will be always reflecting from its boundary 
at $x=b$ with the reflection probability $r^{2}$, no matter how small 
its wavelength is, even {\em at} the classical limit $\lambda=0$.

In other words, although intuitively one would expect the particle moving 
with the energy $E>V$ only to change its speed after passing over the point 
$x=b$, there actually exists a possibility of {\em classical} (the so called 
non-Newtonian, \cite{RS1,RS2}) backward reflections from the sharp potential 
barrier edge.

This curious fact emphasizes an interesting aspect of the connection between
the classical and quantum mechanics, known as the ``Correspondence Principle'',
which was first invoked by Niels Bohr in around 1923. This fundamental principle 
states that classical mechanics can be understood as a limiting case of quantum
mechanics in the so called ``classical limit'', i.e. in case when the motions are
characterized by the actions much larger than the value of the Plank's constant $h$.

At Bohr's time, the fulfillment of such quantum-classical correspondence was viewed
as a natural way to validate meaningfulness and physical consistency of the quantized
analogs of familiar classical systems, such as atoms. Hence the correspondence principle 
is often understood as a naive requirement for the quantum system to reproduce the 
expected classical behavior in the limit $h\rightarrow 0$, whereas in fact, there is no 
such requirement. The appearance of Non-Newtonian scattering events is a curious example 
of a situation when quantum mechanics elucidates a certain implicit aspect of the corresponding 
classical dynamics, bringing up details that could have been easily overlooked.

Similar considerations can prove that non-Newtonian scattering phenomena can happen 
for every potential with sharp edges, as some kind of a reminder of quantum-mechanical 
legacy of classical mechanics. The fact that we never observe such events in every day 
life implies that in reality there are no potentials changes, sharp on the scale of the 
quantum wavelengths of the macroscopic objects.

\section{Non-Newtonian Chaos}

Because of the possibility of such classical non-Newtonian reflections, the 
classical dynamics of a particle in the potential (\ref{pot}) is far from 
trivial. Every time the particle approaches the boundary $x=b$ between the 
two regions in (\ref{pot}), it can be reflected from it with the probability 
$r^{2}$ and transmitted through it with the probability $t^2=1-r^2$. 
As a result, instead of a couple of back and forth oscillations which one 
would naively expect from a particle in the potential (\ref{pot}), the 
actual trajectories of such particle are far more complex. The particle 
can start, say, in the left side $L$ of the well, move to the boundary 
$x=b$, reflect back from it, reflect back from the rigid wall at $x=0$, do 
this several times, then transmit eventually to the right side $R$ of the 
well, where it can also perform a number of oscillations before returning 
to the left side of the well, etc.
Any thinkable sequence of oscillations in the right and in the left sides 
of the potential well (\ref{pot}) represents a possible trajectory of the
particle. 
%%%%%%%%%%%%%%%%%%%%%%%%%%%%%%%%%%%%%%%%%%%%%%%%%%%%%%%%%%%%%%%%%%%%%%
\begin{figure}
\begin{center}
\includegraphics{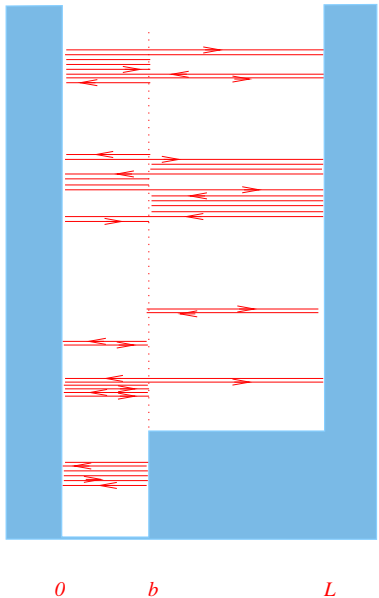}
\caption{\label{Fig.4} 
A storm in a coffee cup - non Newtonian chaos in a one dimensional step 
potential. The lines represent examples of non-Newtonian orbits in the
potential (\ref{pot}). For illustration purposes different trajectories
are shown at different energies, as well as different sections of same 
trajectories are shifted slightly up and down the energy axis, so that 
their parts are visible.}
\end{center}
\end{figure}
%%%%%%%%%%%%%%%%%%%%%%%%%%%%%%%%%%%%%%%%%%%%%%%%%%%%%%%%%%%%%%%%%%%%%% 
It should also be emphasized, that at every reflection or transmission 
(scattering) event, the particle completely looses its memory about the
previous stage of its motion. Given the current position and the momentum 
of the particle, neither its previous evolution nor its state of motion 
after the next collision can be reconstructed. Hence, instead of the 
deterministic evolution we generated a fairly complicated {\em stochastic} 
dynamical process by considering a seemingly simple potential (\ref{pot}).

Obviously, there can be classical orbits of arbitrary length, and the bigger 
is the allowed length (or the period) of the trajectory, the larger are the 
numbers of nontrivial orbits that come into picture. Is there a way to 
enumerate all this variety the orbits? It turns out that the description 
of the general behavior of the orbits in this system can be conveniently 
formalized. Indeed, every orbit can be described by a two-letter code, which 
would simply tell us in what sequence the orbit swings through the left ($L$) 
or the right ($R$) sides of the well, as shown in (Fig.~4).
%%%%%%%%%%%%%%%%%%%%%%%%%%%%%%%%%%%%%%%%%%%%%%%%%%%%%%%%%%%%%%%%%%%%%%
\begin{figure}
\begin{center}
\includegraphics{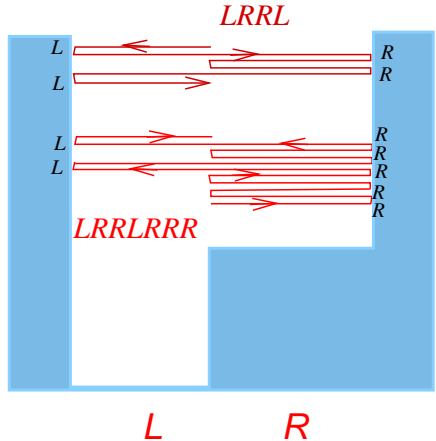}
\caption{\label{Fig.5} A couple of examples of a non-Newtonian orbits in 
the potential (\ref{pot}) marked by their {\em R-L} binary codes. Every left swing 
of the orbit contributes an ``L'' to its code and every right one contributes an ``R''.} 
\end{center}
\end{figure}
%%%%%%%%%%%%%%%%%%%%%%%%%%%%%%%%%%%%%%%%%%%%%%%%%%%%%%%%%%%%%%%%%%%%%% 
Periodic orbits are obviously represented by periodic sequences of symbols. 
Conversely, any periodic sequence of the symbols (for example, $RLRLRLRLRL...$
or just $(RL)$ for short), unambiguously describes a certain periodic orbit  
(note however, that two sequences which can be obtained from one another by 
a cyclical permutation of symbols, correspond to the same periodic orbit).
If the orbit swings $n_{1}$ times in the left side of the well and $n_{2}$
in the right side, its action length is equal 
\begin{equation}
S_{orb}=2n_{1}S_{1}+2n_{2}S_{2},
\label{orb}
\end{equation}
where $S_{1}=L_{1}k$ and $S_{2}=L_{2}\kappa$.

The number of the periodic orbits increases indefinitely. Their shapes (and 
correspondingly their $L-R$ binary codes) become more and more complicated, 
so the classical mechanics in the potential turns out to be surprisingly rich. 
In fact, the number of the geometrically different orbit shapes (or the number 
of the prime periodic orbits, the ones that never retrace themselves) that include 
up to $m$ scattering events grows exponentially as
\begin{equation}
N\approx \frac{e^{0.7 m}}{m}.
\label{rate}
\end{equation}
%%%%%%%%%%%%%%%%%%%%%%%%%%%%%%%%%%%%%%%%%%%%%%%%%%%%%%%%%%%%%%%%%%%%%%
\begin{figure}
\begin{center}
\includegraphics{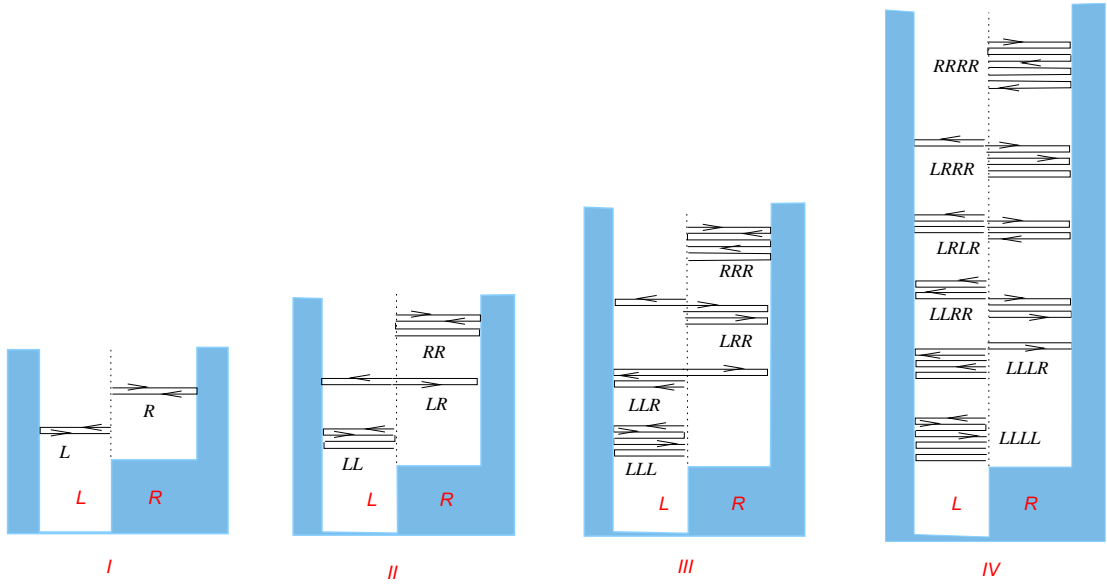}
\caption{\label{Fig.6} Exponential proliferation of the orbits in the potential 
(\ref{pot}). The orbits that undergo 1, 2, 3 and 4 scatterings marked by their 
{\em R-L} binary codes.} 
\end{center}
\end{figure}
%%%%%%%%%%%%%%%%%%%%%%%%%%%%%%%%%%%%%%%%%%%%%%%%%%%%%%%%%%%%%%%%%%%%%% 
Such behavior closely resembles the periodic orbit proliferation scenario that takes 
place in classically nonintegrable chaotic systems. Hence a particle in the potential 
(\ref{pot}) can be viewed as a simple model of a classically chaotic system. This $L-R$
code representation of the orbits is a simple example of symbolic dynamics over a 
partition of the phase space \cite{Cvit,Smale}.

The difference between this and the simple back and forth motion in the 
square well potential is overwhelming, and provides a simple illustration 
to the difference between the dynamical behavior of the integrable and the 
nonintegrable systems, such as e.g. Hydrogen and Helium.
The outburst of complexity that results after adding an extra potential 
step at the bottom of the potential (\ref{well}) is similar to the
situation with the Helium atom. An extra electron that is added to the 
two-body system of the nucleus and the electron in the Hydrogen atom
totally destroys its integrability.

Of course the more complex is the orbit shape, the smaller is the probability 
that this particular orbit will be realized. This ``complexity selection'' can 
be easily understood from the point of view of the quantum mechanics, where the 
propagation of the particle is described by the normalized wave (\ref{wave}). It 
is clear that every reflection or transmission reduces its initial amplitude by 
the amount of the reflection or transmission coefficient.
Hence, it is clear that if a certain prime periodic orbit $p$ (i.e. the one that 
can not be considered a repetition of a shorter orbit) reflects $\sigma(p)$ times 
from the either side of the potential barrier at $x=b$, and transmits $\tau(p)$ 
times through it, its initial amplitude will decrease
\begin{equation}
A_{p}= (-1)^{\chi(p)}r^{\sigma(p)}(1-r^{2})^{\tau(p)/2}
\label{A}
\end{equation}
times \cite{Nova}. Here the factor $(-1)^{\chi(p)}$ keeps track of the sign
changes due to the wall reflections and the right reflections from the boundary.
If the orbit $p$ is traced over itself $\nu$ times, then the corresponding 
amplitude will be $A_{p}^{\nu}$. 
So the quantity (\ref{A}) is a certain quantum (and, since $r$ does not depend
on $\hbar$, also classical) ``weight'' of the orbit. 

\section{Classical Dynamics and Quantum Spectrum}

In the case of the infinite square well potential the orbit bouncing back and 
forth between the walls of (\ref{well}) completely exhausts all the classical 
dynamical possibilities of the particle in the infinite square well. 
This orbit also defines the quantum energy spectrum (\ref{deb}) of the particle 
via the relationship $S_{n}=\pi n$. This is analogous to the general quantization 
procedure of the so called integrable systems, i.e. the ones that have as many 
dynamically conserved quantities (e.g. angular momentum, etc.) as degrees of freedom 
\cite{Einstein,Keller}. These conserved quantities (the action integrals) $I_{i}$, 
$i=1,...,d$, are quantized according to
\begin{eqnarray}
I_{i}=h(n_{i}+\mu_{i}),
\label{ebk}
\end{eqnarray}
where $n_{i}$s are natural numbers, $n_{i}=1,2,...$, and $\mu_{i}$ is a certain 
geometrical constant \cite{Maslov}. The energy of 
these systems is a certain function of the $I_{i}$'s, $E=E(I_{1},...I_{N})$, and 
so the quantization rules (\ref{ebk}) provide immediately the (semiclassical) energy 
spectrum. For example, for a particle moving in the infinite potential well (\ref{well}), 
the magnitude of the momentum $|p|$, and hence the action, $I=2|p|L$, is conserved. It is 
then quantized as $|p|=\hbar\pi n/L$, and yields the quantum energy eigenvalues $E=(\hbar\pi n)^2/2mL^2$.
 
According to this scheme, the ``quantum numbers'' $n_{i}$ are naturally related 
to the classical integrals of motion $I_{i}$.

On the other hand, the classical behavior of the particle in the potential 
(\ref{pot}) is of manifestly nonintegrable type.
The variety of possible dynamical orbits in the potential (\ref{pot}) is much 
richer than in (\ref{well}), and we were able to describe all of them. 
What information is hidden in this variety of the orbits? Is it possible, 
after all, to use this knowledge to obtain the quantum energy spectrum?
One should remember that most of these orbits appeared as some leftovers of 
purely quantum effects - the over-barrier reflections and under-barrier tunnelings.
Do they carry through the complete quantum mechanical information about the particle?

As pointed out above, in the integrable systems there exists a convenient 
handle - the action integrals, which can be ascribed certain discrete values 
via the EBK semiclassical quantization procedure above \cite{Einstein,Keller}.
What if the integrals do not exist? What is there in a nonintegrable system 
that can be used in order to quantize the system semiclassically?

The first answer to this question came in early 70s, when Gutzwiller \cite
{Gutzwiller} showed that a ``handle'' for semiclassical quantization of the 
systems without the sufficient number of the integrals of motion is provided 
by the so-called density of states - a functional that corresponds a $\delta$
-peak to every energy level $E_{n}$ of the quantum spectrum,
\begin{equation}
\rho (E)\equiv \sum_{n=1}^{\infty}\delta\left(E-E_{n}\right).
\label{density}
\end{equation}
Gutzwiller provided a semiclassical expansion for $\rho(E)$ (the so-called 
Gutzwiller trace formula) in terms of the classical quantities related to 
the periodic orbits,
\begin{equation}
\rho (E)\approx\bar{\rho}(E)+\frac{1}{\pi}\mathop{\rm Re}
\sum_{p}T_{p}\sum_{\nu=1}^{\infty}B_{p}^{\nu}\,e^{i\nu S_{p}(E)}.
\label{rho}
\end{equation}
Here $T_{p}$, $S_{p}$ and $B_{p}$ are correspondingly the period, the action and 
a certain weight factor (see below) of the prime periodic orbit $p$ (all classical 
quantities), and $\nu$ is the number of times the orbit $p$ repeats itself. 
The first term $\bar\rho(E)$ represents the non-oscillating part of the density of 
states.

The gist Gutzwiller's trace formula is an (almost miraculous) interference effect, 
produced by the infinity of oscillating terms (one per each periodic orbit) in the 
sum (\ref{rho}). The statement made by (\ref{rho}) is that if the energy $E$ happens 
to coincide with a {\em quantum} energy level $E_n$, then all the terms in the sum 
(\ref{rho}) will interfere constructively and produce a peak, whereas if there are 
no quantum $E_n$'s there, they will interfere destructively and yield $0$. 

This is a truly remarkable connection between the classical and the quantum properties of 
a system. After all, from a formal mathematical perspective, classical characteristics of 
a system should a priori describe only its $\hbar=0$ limit, whereas according to the trace 
formula one can use the classical properties of the system to extract the information about 
its {\em quantum} properties for $\hbar\neq 0$.

Gutzwiller's formula can also be applied to the 1D system of a particle in the step potential 
(\ref{pot}). Moreover, it can be shown \cite{Nova} that for the system (\ref{pot}) the expansion 
(\ref{rho}) is {\em exact}, although usually it provides only an approximate (with semiclassical 
accuracy) representation of $\rho(E)$. Another important characteristic of the spectrum is the 
``spectral staircase'' $N(E)$,
\begin{equation}
N(E)\equiv\int_{0}^{E}\rho(E')dE'=\sum_{n}\Theta\left(E-E_{n}\right),
\label{N}
\end{equation}
where $\Theta(x)$ is Heaviside's theta function,
\begin{equation}
\Theta(x)=\cases{1, &if $x>0$, 
\cr 0 &if $x<0$,\cr}  
\label{heaviside}
\end{equation}
which gives the number of the energy levels on the interval between $0$ and $E$.
$N(E)$ can also be expanded into a periodic orbit series,
\begin{equation}
N(E)=\bar N(E)+\frac{1}{\pi}\mathop{\rm Im}
\sum_{p}\sum_{\nu=1}^{\infty}\frac{A_{p}^{\nu}}{\nu}\,e^{i\nu S_{p}(E)}.
\label{stair}
\end{equation}
What makes the expansion (\ref{stair}) particularly convenient for our use, is that the 
weight factors $A_{p}$ in it are explicitly given by (\ref{A}). The first term of $N(E)$ 
(its non-oscillating part, also known as Weyl's average) is given by
\begin{equation}
\bar N(E)=\frac{1}{\pi}S(E)-\gamma_{0},
\label{weyl}
\end{equation}
where $S(E)$ is the action length (\ref{potaction}) of the classically available region 
of the well, and $\gamma_{0}$ is a small correction term (almost a constant, see below).

The complexity of our task of describing the spectrum in (\ref{pot}) can
already be appreciated from the expansions (\ref{rho}) and (\ref{stair}):
every non-Newtonian orbit that exists in (\ref{pot}) explicitly contributes 
to these expansions.

In addition to the huge number of the periodic orbits, there are many other subtle 
dynamical effects  that contribute to the complexity of the expansions (\ref{rho}) 
and (\ref{stair}). For instance, it was mentioned above that as the energy of the 
particle passes from above $V$ to a level below $V$, the particle becomes classically 
unable to penetrate into the right section of the well, $b<x<L$. However, there exists 
the possibility of quantum tunneling into the region $R$. This is manifested by 
the corresponding momentum becoming imaginary, $\kappa\rightarrow i\varkappa$. 
Hence, the contributions to the orbit actions (\ref{orb}) due to the ``tunneling'' 
through the step parts are also imaginary, $S_{orb}=2S_{1}n_{1}+i2n_{2}S_{2}$ (such orbits
were called ``ghost orbits'' in \cite{Haake}).

That implies that as the energy of the particle changes, the physical characteristics of the
orbits (and therefore of the expansion terms in (\ref{rho})) may change (phenomenon known as
the ``phase space metamorphosis'' \cite{PSM}). However, the formal structure of the Gutzwiller's
formula (at least in the case at hand) remains the same, and so the expansions (\ref{rho}) and 
(\ref{stair}) can be used, with due care, both for $E>V$ and for $E<V$.

\section{Obtaining the spectrum}

Since the right hand side of the expansion (\ref{rho}) can be obtained from 
considering the classical motion of the particle, the expansion (\ref{rho})
provides a clear connection between the dynamical characteristics of the system 
and its spectrum in quantum regime. The energy levels of the quantum system are
obtained according to (\ref{rho}) as the poles (delta spikes) produced by the 
periodic orbit sum \cite{Gutzw,Gutzwiller}. It is important however, that by 
using this approach, one can not tell {\em when} these poles will appear prior 
to performing the summation of the periodic orbit series (\ref{rho}).
Without having any extra information about the system, the only general strategy
for obtaining $E_{n}$'s is to scan the energy axis by summing the series (\ref{rho}) 
for every value of $E$ to find out at what energies the sum produces a $\delta$-peak.

This illustrates the fact that $\rho(E)$ is a {\em global} characteristic of the 
spectrum. Both $\rho(E)$ and its expansion (\ref{rho}) describe the whole spectrum 
at once, rather than the specific individual energy levels, in contrast with the case 
of the {\em integrable systems}, where every action integral $I_{i}$ is quantized 
directly (\ref{ebk}) and {\em separately} from the rest of the degrees of freedom.

Actually, it is possible to extract the information about the individual energy levels 
$E_{n}$ out of $\rho(k)$ without such tedious ``energy axis scanning'' - if only one knows 
(at least approximately) where to look. Indeed, if for instance we would happen to know that 
some energy level $E_{n}$ is the {\em only} level that lies between two points $\hat E_{n}^{+}$ 
and $\hat E_{n}^{-}$ (and hence $\rho(E)$ has only one $\delta$-peak between these points), then 
we could evaluate the integral
\begin{equation}
E_{n}=\int_{\hat E_{n}^{-}}^{\hat E_{n}^{+}}E\rho (E)dE,
\label{root}
\end{equation}
and obtain the value of $E_{n}$.
Since we know the expansion of $\rho(E)$ in terms of the periodic orbits
(the right hand side of (\ref{rho})),
this integration (at least in principle) can be performed.

If our knowledge of a certain system would be so complete that we would 
be able to separate {\em every} energy level $E_{n}$ from its neighbors by
two ``separators'' $\hat E_{n}^{+}$ and $\hat E_{n}^{-}$, 
$\hat E_{n}^{+}>E_{n}>\hat E_{n}^{-}$, then we could use (\ref{root}) to 
obtain an exact numerical value of every energy level in the spectrum of our
problem out of $\rho(E)$. 
Apparently, for the purpose of separating one level from another, it is enough 
to assume that $\hat E_{n-1}^{+}=\hat E_{n}^{-}$, so basically we are looking 
for 
a sequence of points $\hat E_{n}$ that interweaves the sequence of the energy 
levels and separates one energy level from another \cite{Opus,Prima,Sutra,Stanza}. 
In other words, in order to be able to obtain {\em quantitatively} any energy 
level $E_{n}$ out of $\rho(E)$, we need to establish a partition of the $E$-axis 
into 
the intervals $\left[\hat E_{n-1},\hat E_{n}\right]$, each one of which 
contains exactly one energy level. 

Note, that even if that could be achieved, one would still have just an 
algorithmic recipe for evaluating $E_{n}$'s rather than a {\em formula} of 
the type $E_{n}=...$ . In order to get such a {\em formula}, we would need 
to find {\em a global function} $\hat E(n)$ that depends explicitly on $n$ 
and produces all the separating points in their natural sequence {\em as a 
function of their index},
\begin{eqnarray}
\hat E_{n}=\hat E(n).
\label{separators}
\end{eqnarray}
Once such a functional dependence of the separators on their index is 
established,
the integral (\ref{root}) would turn the index $n$ into a {\em quantum number} 
and 
give us a complete solution to the spectral problem in the form  $E_{n}=...$ .

The problem is that usually this is not an easy task. In order to establish a 
partition of the energy axis into the separating intervals $\left[\hat E_{n-1},
\hat E_{n}\right]$ one needs to have some extra information about the behavior 
of the spectrum. Luckily, such information can be obtained for the particle in 
the potential (\ref{pot}).

\section{Spectral equation}

In order to get this information, let us examine closely the spectral equations 
(\ref{specabove}) and (\ref{specbelow}). Since the quantity that we are after, the 
classical action $S(E)$ defined by (\ref{potaction}) is in the arguments of the sines
in the left hand sides of the equations (\ref{specabove}) and (\ref{specbelow}), we
can formally invert them to obtain
\begin{eqnarray}
S=\pi\left(n-\frac{1}{2}\right)-\cases{(-1)^{n}
\left(\arcsin\left(r\sin\left(S_{1}-S_{2}\right)+\frac{\pi}{2}\right)\right),
&if $E>V$, \cr
\arctan\left(\varkappa\left(1+e^{-2S_{2}}\right)/k\left(1-e^{-2S_{2}}\right)\right), 
&if $E<V$, \cr}
\label{inverse}
\end{eqnarray}
where $n=1,2,... \,$.
This form of the spectral equation is particularly important for two reasons. First, 
we are getting an index $n$ which (one would assume) numbers the solutions $S_{n}$ - 
the discrete quantum values of action. 

One can notice an interesting resemblance between the way the formulae (\ref{inverse}) 
begin, $S=\pi\left(n-\frac{1}{2}\right)+...$, and the formula (\ref{ebk}). One can 
speculate therefore that the first term, $\pi\left(n-\frac{1}{2}\right)$, expresses 
the regular part of the spectrum, and the second terms, which actually make the equation 
(\ref{inverse}) transcendental, are introducing the irregularities into the spectrum, 
which are due to the classical non-integrability of the potential well (\ref{pot}).

This in fact turns out to be a very important observation. Is the spectrum defined 
by (\ref{inverse}) more ``regular'' or ``irregular''? Which one of these two terms 
contributes more to the solution, the regular $\pi\left(n-\frac{1}{2}\right)$ part 
or the irregular second term? Surprisingly, it turns out that in terms of the magnitudes 
the regular part wins. Indeed, it is well known that the inverse trigonometric functions 
are bounded, $-\frac{\pi}{2}\leq\arcsin(x), \arctan(x)\leq\frac{\pi}{2}$. 
On the other hand, when the index $n$ in the equation (\ref{inverse}) changes by $1$ 
as we go from one level to another, the ``regular term'' increases by $\pi$, which is 
generically (that is almost always) {\em more} than what the arcsine or arctangent can 
provide. 

Therefore, it is the regular first term $\pi\left(n-\frac{1}{2}\right)$ in (\ref{inverse}) 
that contributes the most to the solution. If the step at the bottom of the well would 
disappear, so would the ``transcendental'' terms in (\ref{inverse}), and a pure regular 
(periodic) spectrum (\ref{deb}) for $S$ would be recovered. Hence the second term in 
(\ref{inverse}) can be regarded as the one responsible for the chaos induced spectral 
``fluctuations''.
 
The second important consequence of writing the equations (\ref{specabove}) and (\ref{specbelow}) 
in the inverted form (\ref{inverse}), is that it allows one to obtain the set of separating points 
discussed in the previous section, and hence to implement the solution to the spectral problem for 
the potential (\ref{pot}).

Indeed, since the irregular terms in (\ref{inverse}) are different from $0$, the values 
$\hat S_{n}=\pi(n-1/2)$ {\em themselves} are never the solutions to (\ref{inverse}). 
On the other hand, the difference between $\hat S_{n}$ and the roots $S_{n}$, which is due 
to the second term, is smaller than $\pi$, and as a result the roots $S_{n}$ of (\ref{inverse}) 
will be {\em locked inside of the intervals}
\begin{equation}
\pi\left(n-\frac{1}{2}\right)<S_{n}<\pi\left(n+\frac{1}{2}\right).
\label{range}
\end{equation}
Thus, the first discrete quantum action value will be locked between $\pi/2$ and $3\pi/2$, 
the second one between $3\pi/2$ and $5\pi/2$, and so on. In other words, we run immediately 
into the separating points for the quantum values of action, 
\begin{equation}
\hat S_{n}=\pi(n-1/2), \ \ n=1,2,...\, ,
\label{actionseps}
\end{equation}
or the {\em action separators}. 
These separators can be used to extract the energy separators $\hat E_{n}$ from 
$S(\hat E_{n})=\pi(n-1/2)$, which then can be used to obtain the energy spectrum via 
(\ref{root}).
%%%%%%%%%%%%%%%%%%%%%%%%%%%%%%%%%%%%%%%%%%%%%%%%%%%%%%%%%%%%%%%%%%%%%%
\begin{figure}
\begin{center}
\includegraphics{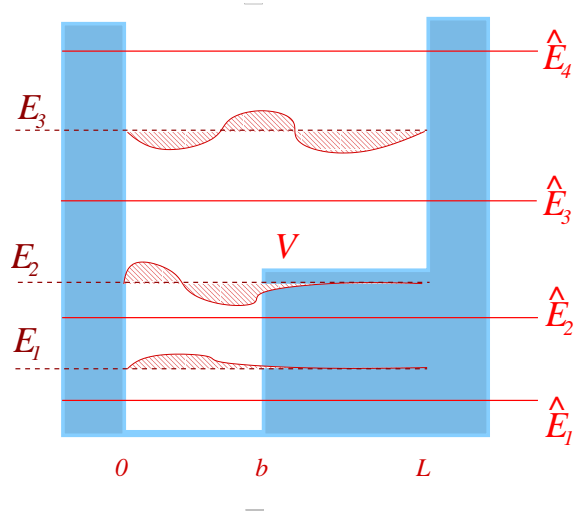}
\caption{\label{Fig.7} The energy separators $\hat E_{n}$ (their values on the energy scale are shown 
by solid lines) define the upper and the lower bounds for the quantum energy levels $E_n$, shown by the
dashed lines.} 
\end{center}
\end{figure}
%%%%%%%%%%%%%%%%%%%%%%%%%%%%%%%%%%%%%%%%%%%%%%%%%%%%%%%%%%%%%%%%%%%%%% 
Alternatively, rather than solving the equation $S(\hat E_{n})=\pi(n-1/2)$ for the $\hat E_n$'s in order 
to use them in the formula (\ref{root}), one could introduce a ``density of the action states'', $\rho(S)$, 
defined as $\rho(S)dS=\rho(E)dE$, 
\begin{equation}
\rho (S)\equiv \sum_{n=1}^{\infty}\delta\left(S-S_{n}\right)\frac{dE}{dS},
\label{actiondensity}
\end{equation}
to obtain the actual quantum levels of action first and then to extract the quantum energy levels from them. 
This would completely parallel the $\rho(E)$ approach outlined in the previous section. A simple change of 
variables will produce the periodic orbit expansion for $\rho(S)$. Hence it is possible, in accordance with 
the ideas outlined above, to obtain the discrete quantum action values $S_{n}$ via
\begin{equation}
S_{n}=\int_{\pi(n-1/2)}^{\pi(n+1/2)}S\rho (S)dS=S\,N(S)\Big |_{\pi(n-1/2)}^{\pi
(n+1/2)}-
\int_{\pi(n-1/2)}^{\pi(n+1/2)}N(S)\,dS,
\label{rootS}
\end{equation}
where we used the identity 
\begin{equation}
\rho (S)=\frac{dN(S)}{dS}.
\label{rhoN}
\end{equation}
Using the relationship $N(\hat S_{n})=n$ (there are exactly $n$ roots $s_{n}$ below the $n$th separator)
and the expansion (\ref{stair}), we get
\begin{eqnarray}
S_{n}=2\pi n-\frac{\pi}{2}-\int_{\pi(n-1/2)}^{\pi(n+1/2)}\bar N(S)dS-
\frac{1}{\pi}\mathop{\rm Im}\sum_{p,\nu}
\int_{\pi(n-1/2)}^{\pi(n+1/2)}
\frac{A_{p}^{\nu}}{\nu}e^{i\nu S_{p}}dS,
\label{Sn}
\end{eqnarray}
where the weight factors $A_{p}$ are given by (\ref{A}).

Note, that (\ref{deb}) is exactly the first term of the expansion (\ref{Sn}). 
Hence the oscillatory terms in (\ref{Sn}), with their amplitudes proportional to the 
non-Newtonian reflection amplitudes, are indeed due to the ``nonintegrability'' of (\ref{pot}).

Since all of the quantities on the right-hand side of (\ref{Sn}) are known, this formula provides 
an explicit representation of the discrete quantum values of the action functional in terms of the 
geometric and dynamical characteristics of the potential (\ref{pot}). Formula (\ref{Sn}) allows the 
computation of the action corresponding to every quantum level {\em individually}, {\it explicitly} 
and {\it exactly} in terms of the classical parameters, indexed by its ``quantum number'' $n$.

It should be mentioned however, that the quantum number $n$ in (\ref{deb}) and in (\ref{Sn}) have 
rather different origins. Formula (\ref{Sn}) does not have such a direct geometrical interpretation 
as (\ref{deb}). While in (\ref{deb}) the number $n$ is the number of waves that fit on the trajectory, 
in the case of the step potential (\ref{pot}) $n$ is the index of the cell (\ref{range}) that contains 
the corresponding action value $S_{n}$.

So once again, we arrive at defining the quantum energy spectrum via a discrete set of allowed values 
of the action functional, $S_{n}=S(E_{n})$. This relationship (which is yet to be presented explicitly) 
can be viewed as a direct generalization of (\ref{deb}).

Of course, for using the periodic orbit expansions in practice, one needs to know 
how to truncate the series (\ref{rho}) or (\ref{Sn}) to obtain finite order approximations 
to $S_{n}$. Since usually the periodic orbit series are not absolutely convergent, the order 
in which the expansion terms are incorporated into the sum (\ref{Sn}) is important. 
It turns out \cite{Opus,Prima,Sutra,Stanza}, that the correct way to obtain 
$m$th approximation to the exact value of $S_{n}$ is to include into the 
sum (\ref{Sn}) all the trajectories that reach the point $x=b$ $m$ times or 
less.

\section{Overview and an example}

Let us summarize the steps for obtaining the $m$th correction to $S_{n}$ using the 
semiclassical periodic orbit expansion technique.
\goodbreak
\begin{enumerate}
\item  Write down all the $m$-letter sequences (words) $w_{m}$. For example, 
there are 2 words, $(L)$ and $(R)$, for $m=1$, 4 words $(LL)$, $(LR)$, $(RL)$ 
and $(RR)$ for $m=2$, 8 words $(LLL)$, $(LLR)$, $(LRL)$, $(RLL)$,  $(RLR)$, 
$(RRL)$, $(LRR)$ and $(RRR)$ for $m=3$ and so on.
(To do the same thing geometrically - draw all possible periodic orbits similar 
to the ones shown in Fig.~5 that include exactly $m$ transmissions and reflections).
\item Find which $w$'s are cyclic permutations of one another - all these
sequences represent the same orbit, so pick one and discard its replicas.
From the above examples, both $m=1$ orbits will remain, for $m=2$ we may
keep
$(LL)$, $(LR)$ and $(RR)$, for $m=3$ we keep 
$(LLL)$, $(LLR)$, $(RLR)$ and $(RRR)$.  
(Geometrically: find out which loops are representing the same sequence of 
left and right swings. Pick one and discard its replicas).
\item  Count the number $n_{L}$ of $L$'s and $n_{R}$ of $R$'s in 
$w$ and find the action $S_{w}$ of the orbit $w$ according to 
\begin{equation}
S_{w}=2n_{R}S_{1}+2n_{L}S_{2}.
\end{equation}
Remember that if $E<V$ then $S_2=i\sqrt{V-E}L_{2}$.
\item Assuming that the first symbol in $w$ cyclically follows the last one, 
find out which of the remaining sequences $w$ are prime sequences and which 
ones are repetitions of a shorter code. Find the repetition number $\nu_w$ for 
each orbit (use $\nu_w=1$ if the orbit is prime).
For example, for $m=2$ the orbit $\nu_{LR}=1$, $\nu_{RR}=\nu_{LL}=2$, and
for $m=3$ $\nu_{LLL}=\nu_{RRR}=3$, $\nu_{LLR}=\nu_{RLR}=1$.  
(Geometrically, find out how many times each orbit traverses over itself).
\item Scan each word $w_{m}$ (again, the first symbol of $w$ follows the last 
one) and write down the weight $A_{w}$ according to the substitutions
\begin{equation}
LR\rightarrow t,\ \
RL\rightarrow t,\ \
LL\rightarrow r,\ \
RR\rightarrow -r,
\end{equation}
plus each wall reflection contributes a sign change.
For example, for $m=1$, $A_{L}=-r$, $A_{R}=r$, for $m=2$ $A_{LR}=t^{2}$, 
$A_{RR}=A_{LL}=r^{2}$, and for $m=3$ we have $A_{LLL}=-r^{3}$, $A_{RRR}=r^{3}$, 
$A_{LLR}=-rt^{2}$, $A_{RLR}=rt^{2}$. 
(Geometrically: assign a factor $t$ to every transmission, a factor $-1$ to 
every wall reflection, a factor $r$ to every left side reflection and a factor 
$-r$ to every right side reflection).
\item For every orbit, find out the corresponding term in the expansion (\ref{Sn}),
\begin{equation}
\int_{\pi(n-1/2)}^{\pi (n+1/2)}\frac{A_{w}}{\nu_{w}}e^{iS_{w}(S)}dS,
\end{equation}
and integrate.
\end{enumerate}
If all the words up the length $m$ are considered, the result will produce 
$m$th approximation to the exact discrete sequence of actions. 

All the non-Newtonian orbits described in the previous sections contribute to the
expansion of each eigenvalue $S_{n}$. One can now appreciate the complexity of solving 
analytically the equations (\ref{specabove}) and (\ref{specbelow}) compared to solving 
the simple equation (\ref{wellspec}), by comparing the complexities of the corresponding 
classical dynamics.

\subsection{An example}

As an example, let us compute the first three approximations to a a few momentum eigenvalues,
using the orbits that include up to 3 scattering events, outlined in the summary and shown in Fig.~6.
%%%%%%%%%%%%%%%%%%%%%%%%%%%%%%%%%%%%%%%%%%%%%%%%%%%%%%%%%%%%%%%%%%%%%%
\begin{figure}
\begin{center}
\includegraphics{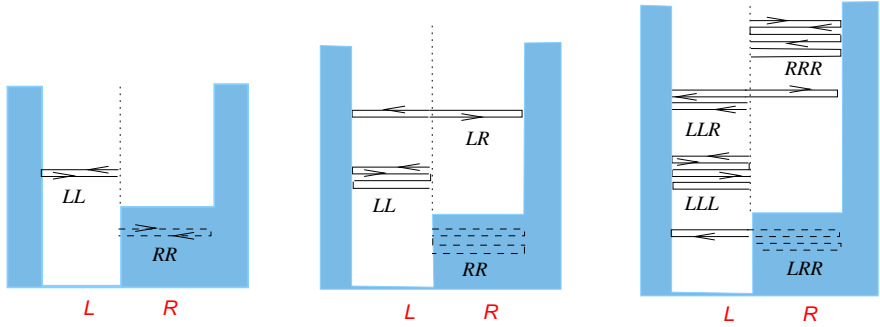}
\caption{\label{Fig.8} 
The orbits required for evaluating the first three periodic orbit expansion approximations 
to the $S_{n}$. Since the evaluation of the low levels which correspond to the energy $E<V$ 
will require ghost orbits, a few ghost orbits are also illustrated. The tunneling parts of 
the ghost orbits are shown with dashed lines.}
\end{center}
\end{figure}
%%%%%%%%%%%%%%%%%%%%%%%%%%%%%%%%%%%%%%%%%%%%%%%%%%%%%%%%%%%%%%%%%%%%%% 
We have according to (\ref{Sn}) that the first three corrections will be:
\begin{eqnarray}
S_{n}&=&2\pi n-\frac{\pi}{2}-\int_{\pi(n-1/2)}^{\pi(n+1/2)}\bar N(S)dS \cr
&-&\mathop{\rm Im}\int_{\pi(n-1/2)}^{\pi (n+1/2)}
\left(re^{i2S_{1}k}-e^{i2S_{2}k}r\right)dS \cr
&-&\mathop{\rm Im}\int_{\pi(n-1/2)}^{\pi (n+1/2)}\left(
\frac{1}{2}r^{2}e^{4iS_{1}}+t^{2}e^{i2S_{1}+i2S_{2}}+\frac{1}{2}e^{4iS_{2}}r^{2}\right)dS \cr
&-&
\mathop{\rm Im}\int_{\pi(n-1/2)}^{\pi (n+1/2)}\left(\frac{1}{3}r^{3}e^{6iS_{1}}+t^{2}re^{i\left(2S_{2}+4S_{1}\right)}
-t^{2}\allowbreak re^{i\left(2S_{1}+4S_{2}\right) }
-\frac{1}{3}r^{3}e^{6iS_{2}}\right)dS.
\label{3terms}
\end{eqnarray}
In order to use the formula (\ref{3terms}), one needs to specify the parameters of the potential
well (\ref{pot}). In particular, it is a specific value of $V$ that determines at what energy the 
periodic orbits will acquire tunneling parts. 
For simplicity, let us consider the case of symmetrical well, when $L_{1}=L_{2}=1$.
Then the action $S$ defined by (\ref{potaction}) will be
\begin{equation}
S(E)=\cases{k+\kappa, &if $E>V$, 
\cr 
k&if $E<V$,\cr}  
\label{exaction}
\end{equation}
in terms of which
\begin{equation}
\cases{
r=\frac{V}{S^{2}}, \ \ t^2=\left(1-\frac{V^2}{S^4}\right), &if $E>V$,
\cr
r=e^{2i\pi\gamma_{0}}, \ \ \gamma_{0}=\frac{1}{\pi}\arctan\frac{\sqrt{V-S^{2}}}{S}, 
\ \ t^2=\left(1-e^{2i\pi\gamma_{0}}\right) &if $E<V$.
\cr}
\label{actr}
\end{equation}
The phase $\gamma_{0}$ (which also appears in the Weyl's average (\ref{weyl})) is $\gamma_{0}=1/2$
for $E>V$.

Next, let us pick a certain value for the potential step hight, say $V=25$. This will set 
the $E_{crit}=25$ and $k_{crit}=5$ to be the ``critical values'' of energy and the momentum - that is if 
we will be looking for an energy level below $E=V=25$, then the tunneling effects (ghost orbits, 
the restructuring of the periodic orbit sum (\ref{rho})) have to be taken into account. 
Correspondingly, the critical value for the action will be $S_{crit}=k_{crit}L_{1}=5$, and so one should
use different parts of formulae (\ref{exaction}) to extract the $E_n$ or $k_n$ out of $S_n$, depending on 
whether a particular $s_{n}$ is smaller or greater than $5$. 

After the height of the potential step has been chosen, certain general statements about the behavior of 
the quantum $S_n$ levels can be made. For the specific case $V=25$, the first and the second separators, 
$\hat S_{1}=\frac{\pi}{2}$ and $\hat S_{2}=\frac{3\pi}{2}$, will be smaller than $S_{crit}$, and so the 
level contained between them will correspond to the energy $E_{1}<V$. The third action separator is bigger 
than $k_{crit}$, $\hat S_{3}=5\pi/2>5$, so we will be in a better position to judge about the location of 
the second level $\hat S_{2}< S_{2}<\hat S_{3}$ after computing the corrections (\ref{3terms}). All the other 
levels $S_{n}$, for $n>2$, have energies higher than $V$.

So let us find the discrete quantum action values $S_{1}$, $S_{2}$ and (for example) $S_{17}$, and 
the corresponding quantum momentum eigenvalues, $k_{1}$, $k_{2}$ and $k_{17}$. Using the expressions 
(\ref{3terms}), (\ref{exaction}), (\ref{actr}) and integrating from $\pi/2$ to $3\pi/2$, from $3\pi/2$ 
to $5\pi/2$ and from $33\pi/2$ to $35\pi/2$ (using some numerical integration software is highly recommended) 
we find the values shown in Table 1.
\bigskip 
\begin{table}[ht]
\begin{tabular}{c|c|c|c|cc|c}
root & $m=0$ & $m=0$ & $m=2$ & $m=3$ &  & exact\\ 
%&  &  &  &  &  &  \\
\hline 
$s_{1}$ \ & 2.4354 \ & \ 2.6198 \ & \ 2.6173 \ & \ 2.605 \ & \  \ & \ 2.5958787201295728 \\ 
\hline
$s_{2}$ \ & \ 6.1601 \ & \ 5.5789 \ & \ 5.2366 \ & \ 5.1434 \ & \  \ & \ 4.9455316914381690 \\ 
\hline
$s_{17}$ \ & \ 53.4071 \ & \ 53.405 \ & \ 53.404 \ & \ 53.406 \ & \  \ & \ 53.403119615030526
\end{tabular}
\bigskip
\caption{
The results of the periodic orbit expansion approximations to the exact values of $s_{n}$. 
The first $m=0$ column gives the Weyl's $s_{n}^{(m=0)}$ estimate for the roots of the spectral 
equation, and the following columns show the corrections to it due to the $m=1$, $m=2$ and $m=3$ 
code length orbits. Note that the third periodic orbit expansion approximation to $s_{2}$ comes out
higher than the critical value, $s_{2}^{(m=3)}>s_{crit}=5$, whereas the actual root $s_{2}$ is smaller 
than $s_{crit}$. Hence more periodic orbit expansion terms are needed to capture the correct qualitative
behavior of $s_{2}$.}
\label{table}
\end{table}

Note that in the second case the integration in (\ref{3terms}) should be split in two parts,
\begin{equation}
\int_{3\pi/2}^{5\pi/2}N(S)dS=
\int_{3\pi/2}^{5}N^{(E<S)}(S)dS+\int_{5}^{5\pi/2}N^{(E>V)}(S)dS,
\label{split}
\end{equation}
in order to follow the changes in the structure of the expansion terms described by (\ref{exaction}) and 
(\ref{actr}). Note also that the column to column changes are more significant for the second root - this 
indicates that the ``spectral fluctuations'' for this root are high due to the orbit metamorphosis that takes 
place at $E=V$, inside of the integration interval $\hat E_{n-1}<E_{n}<\hat E_{n}$. This also indicates that 
more periodic orbit expansion terms are needed to get the more precise position of $s_{2}$ with respect to $V$.

After the allowed values  $S_{n}$ for the action are found, the quantum levels of the momentum can be obtained 
by inverting the relationship (\ref{exaction}),
\begin{equation}
k=\cases{\frac{s^{2}-V}{2s}, &if $s_{n}<V$, 
\cr 
s&if $s_{n}<V$.\cr}  
\label{extract}
\end{equation}
According to (\ref{table}) one has from (\ref{extract}):
\begin{eqnarray}
k_{1}= 2.605, \ \ k_{2}= 5.0020, \ \ k_{17}=26.4689.\cr
\label{kcorrections}
\end{eqnarray}
This demonstrates how the periodic orbit expansion quantization rule (\ref{Sn}) gives the explicit solution 
to our spectral problem.

\section{A simplification - scaling potential}

It is possible to simplify significantly the result by assuming that the hight 
of the potential step $V$ also grows ({\em scales}) with energy, $V=\lambda\, E$. 
This assumption is not as artificial as it may seem. Scaling (not necessarily in 
the $V=\lambda\, E$ form) is a common phenomenon that occurs in a variety of familiar 
physical systems \cite{PSM,Tom}, such as, e.g. the Hydrogen atom in a uniform magnetic 
field (also a nonintegrable system). Basically, the scaling eliminates the geometrical 
restructuring of the trajectories - i.e. in our case the transitions from classically 
allowed to ``ghost'' to orbits.

As soon as a certain value of the scaling coefficient $\lambda$ has been chosen, the 
orbits will either always hover above the potential step ($\lambda<1$) or sit below it 
($\lambda>1$). The assumption $V=\lambda\, E$ also implies that the reflection coefficient 
$r_{sc}$ (and therefore the weight factors (\ref{A})) do not depend on energy,
\begin{equation}
r_{sc}=\frac{1-\beta}{1+\beta},
\label{rsc}
\end{equation}
where $\beta=\sqrt{1-\lambda}$. In addition, the action lengths of the two parts 
of the potential (\ref{pot}) become simply proportional to $k$, $S_{1}=L_{1}k$ and 
$S_{2}=L_{2}\beta k$. The spectral equation for $\lambda<1$ is now
\begin{equation}
\sin(\Omega_{0}k)=r_{sc}\sin(\omega k)
\label{scspectr}
\end{equation}
where $\Omega_{0}=L_{1}+\beta L_{2}$ and $\omega=L_{1}-\beta L_{2}$ are constant
frequencies, $\Omega_{0}>\omega$, and the roots of (\ref{scspectr}) are separated 
from one another by a periodic sequence of separators
\begin{equation}
\hat k_{n}=\frac{\pi}{\Omega_{0}}\left(n+\frac{1}{2}\right).
\label{scseps}
\end{equation}
For a given $\lambda$, the coefficients $r_{sc}$, $\Omega_{0}$, $\omega$ and the separators 
(\ref{scseps}) do not depend on energy anymore. The spectral equation (\ref{scspectr}) doesn't 
change its functional form, and the tunneling (ghost) trajectories never appear in the system. 
This clearly illustrates the idea of the ``structural freezing'' of the dynamics due to the
scaling.

With so many characteristics of the system becoming constant, the evaluation of the 
$k_n$ greatly simplifies and the integration in (\ref{Sn}) can be carried out explicitly 
\cite{Opus,Prima,Sutra,Stanza}. 

Let us assume that energy scales above the potential step, $\lambda<1$. The exact eigenvalues 
of the momentum in this case are
\begin{eqnarray}
k_{n}=\frac{\pi}{\Omega_{0}}n-\frac{2}{\pi}\sum_{p}
\frac{1}{S_{p}^{0}}\sum_{\nu=1}^{\infty}
\frac{A_{p}^{\nu}}{\nu^{2}}\sin\left(\frac{\nu\omega_{p}}{2}\right)\,
\sin\left(\nu\omega_{p}n\right),
\label{kn}
\end{eqnarray}
where $S_{p}^{0}=S_{p}/k$, $\omega_{p}=\pi S_{p}^{0}/S_0$, and the weight factors $A_{p}$ are 
given by (\ref{A}) in which $r=r_{sc}$. 
This expression clarifies the overall structure of the periodic orbit expansions for $k_n$.

\section{Discussion}

We have studied an elementary example of a classically nonintegrable 
(quantum stochastic) system in the potential (\ref{pot}). Despite the 
simplicity of the setup, the classical dynamics of the particle in the 
potential (\ref{pot}) turns out to be extremely complicated, and this 
complexity is manifested in its spectral properties in quantum regime.
Surprisingly, in studying this most elementary example, one runs into 
essentially all the dynamical and physical effects (integrability versus
nonintegrability, the exponential proliferation of the periodic orbits in
the nonintegrable case, periodic orbit expansions, the use of symbolic
dynamics, phase space metamorphosis, tunneling, ray splitting, etc.) that 
appear in semiclassical analysis of more realistic physical systems.
Usually these phenomena are investigated via a rather involved mathematical 
apparatus, whereas in our example they naturally come into play and can be 
analyzed by elementary means. Due to its illustrative simplicity, this 
and similar \cite{Anima,Saga,Fabula} systems can be thought of as the ``Harmonic
oscillators'' of quantum chaos. On the other hand, this problem is rich enough to 
illustrate the essence of the difficulties associated with the semiclassical quantization 
of chaotic dynamical systems, such as Helium atom. 

%%%%%%%%%%%%%%%%%%%%%%%%%%%%%%%%%%%%%%%%%%%%%%%%%%%%%%%%%%%%%%%%%%%%%%
\begin{figure}
\begin{center}
\includegraphics{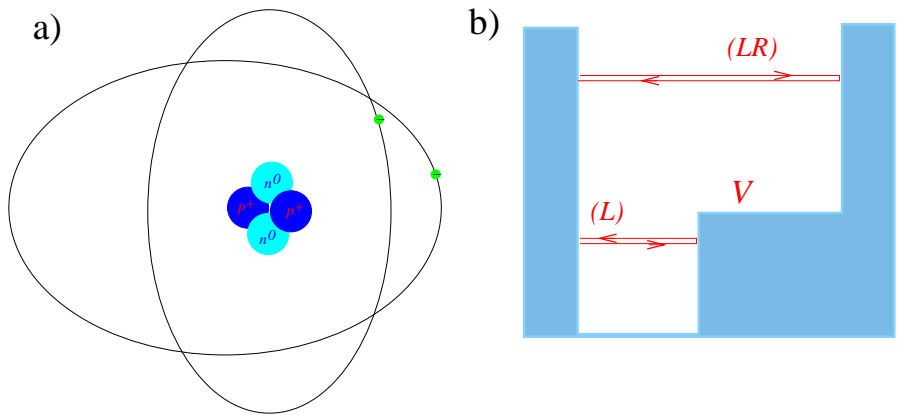}
\caption{\label{Fig.9} 
The familiar schematic picture of the Helium atom (Fig.~7a), which was also used as 
the physical model in 1920s, is analogous to using only Newtonian trajectories in the 
potential (\ref{pot}), shown in Fig.~7b. This scheme completely ignores the actual 
classical dynamical complexity. Compare the Fig.~7(b) to Fig.~3 and Fig.~5.}
\end{center}
\end{figure}
%%%%%%%%%%%%%%%%%%%%%%%%%%%%%%%%%%%%%%%%%%%%%%%%%%%%%%%%%%%%%%%%%%%%%% 
As mentioned above, the early attempts to quantize the Helium atom failed because 
the qualitative difference in the dynamical complexity between Hydrogen and Helium 
were overlooked. The attempts to quantize a chaotic system within the framework of 
Bohr-Sommerfeld or EBK quantization theory using only a few integrable-like trajectories
can be compared to considering just the two ``naive'' (Newtonian) classical trajectories 
in the potential (\ref{pot}), $(L)$ for $E<V$ and $(LR)$ for $E>V$. 
From the structure of the exact result (\ref{Sn}) it is clear that such consideration 
would produce only very approximate results, far from the real complexity of the problem. 

Formula (\ref{Sn}) represents the ``modification'' of the EBK quantization condition (\ref{deb}) 
mentioned in the Van Vleck citation above, for the simple case of the potential (\ref{pot}), in 
an explicit and self-contained form. 
It would be natural to expect that obtaining the semiclassical 
spectrum in the form $E_{n}=...$ for more complicated systems such as Helium atom should be a much 
more difficult task. However, this result creates an interesting precedent that may indicate new 
directions in the semiclassical quantization theory and related fields.
%%%%%%%%%%%%%%%%%%%%%%%%%%%%%%%%%%%%%%%%%%%%%%%%%%%%%%%%%%%%%%%%%%%%%%%%
\begin{acknowledgments}
I would like to thank Reinhold Bl\"umel and Roderick Jensen from Wesleyan 
University with whom the work on the exact spectral formulae has began. I 
would also like to thank Mark Kvale from the UCSF Keck Center for reading 
the manuscript and making a number of useful suggestions.

Work at UCSF was supported in part by the Sloan and Swartz foundations.
\end{acknowledgments}
%%%%%%%%%%%%%%%%%%%%%%%%%%%%%%%%%%%%%%%%%%%%%%%%%%%%%%%%%%%%%%%%%%%%%%%%

%\pagebreak

%\begin{references}

%\end{references}

\end{document}